\documentclass[%
reprint,
superscriptaddress,
 amsmath,amssymb,
 pra,
]{revtex4-2}

\usepackage{lipsum} 

\usepackage[normalem]{ulem}

\usepackage{graphicx}
\usepackage{dcolumn}
\usepackage{bm}

\usepackage{xcolor}
\usepackage{braket}

\usepackage{color}
\definecolor{DarkGreen}{rgb}{0, 0.6, 0.3}

\begin{document}

\preprint{APS/123-QED}

\title{Preservation and destruction of the purity of two-photon states in the
interaction with a nanoscatterer}

\author{\'{A}lvaro Nodar}
\affiliation{Centro de Física de Materiales (CFM), CSIC-UPV/EHU, Paseo Manuel de Lardizabal 5, 20018 Donostia-San Sebastián, Spain}
\author{Ruben Esteban}
\affiliation{Centro de Física de Materiales (CFM), CSIC-UPV/EHU, Paseo Manuel de Lardizabal 5, 20018 Donostia-San Sebastián, Spain}
\affiliation{Donostia International Physics Center (DIPC), Paseo Manuel de Lardizabal 4, 20018 Donostia-San Sebastián, Spain}
\author{Carlos Maciel-Escudero}
\affiliation{Centro de Física de Materiales (CFM), CSIC-UPV/EHU, Paseo Manuel de Lardizabal 5, 20018 Donostia-San Sebastián, Spain}
\affiliation{CIC NanoGUNE BRTA and Department of Electricity and Electronics, UPV/EHU, Tolosa Hiribidea 76, 20018 Donostia-San Sebastián, Spain}
\author{Jon Lasa-Alonso}
\affiliation{Centro de Física de Materiales (CFM), CSIC-UPV/EHU, Paseo Manuel de Lardizabal 5, 20018 Donostia-San Sebastián, Spain}
\affiliation{Donostia International Physics Center (DIPC), Paseo Manuel de Lardizabal 4, 20018 Donostia-San Sebastián, Spain}
\author{Javier Aizpurua}
\affiliation{Centro de Física de Materiales (CFM), CSIC-UPV/EHU, Paseo Manuel de Lardizabal 5, 20018 Donostia-San Sebastián, Spain}
\affiliation{Donostia International Physics Center (DIPC), Paseo Manuel de Lardizabal 4, 20018 Donostia-San Sebastián, Spain}
\author{Gabriel Molina-Terriza}
\affiliation{Centro de Física de Materiales (CFM), CSIC-UPV/EHU, Paseo Manuel de Lardizabal 5, 20018 Donostia-San Sebastián, Spain}
\affiliation{Donostia International Physics Center (DIPC), Paseo Manuel de Lardizabal 4, 20018 Donostia-San Sebastián, Spain}
\affiliation{IKERBASQUE, Basque Foundation for Science, María Díaz de Haro 3, 48013 Bilbao, Spain.}
\date{\today}

\begin{abstract}
The optical resonances supported by nanostructures offer the possibility to enhance the interaction between matter and the quantum states of light. In this work, we provide a framework to study the scattering of quantum states of light with information encoded in their helicity by a nanostructure. We analyze the purity of the scattered output quantum state, and we find that the purity of the incident state can be lost, when it interacts with the optical resonances of the nanostructure. To explain the loss of quantum purity, we develop a physical picture based on time delays and frequency shifts between the output two-photon modes. The framework and analysis proposed in this work establishes a tool to address the interaction between quantum light and nanoenvironments.
\end{abstract}

\maketitle


Quantum entanglement is a fragile resource required for most quantum applications. There have been extensive studies to quantify and exploit the degree of entanglement of different systems, and also study the loss of entanglement and purity through decoherence \cite{zurek2003decoherence, kim2012protecting}. One of the benefits of encoding quantum information in optical states is that they are very resilient to decoherence, while at the same time, by their very nature, photons are excellent information carriers for quantum communication protocols \cite{krenn2014communication, krenn2016quantum, aspelmeyer2003long}. However, photons do not interact strongly with material particles and structures, which limits the possibilities of processing photonic quantum information \cite{ladd2010quantum, nielsen2002quantum}. 

Several techniques are being developed in order to enhance photon interactions, such as quantum optomechanical systems, optical metamaterials, high-density gases, slow-light materials and several others. Engineering nanophotonic nanostructures for quantum information processing offers the possibility to, not only enhancing light-matter interactions, but also manipulating light in devices with a footprint of the order of the wavelength. While the use of nanostructures and the study of their optical resonances to enhance the classical interaction of light and matter has a long tradition \cite{IntroCl1, IntroCl2, IntroCl3, IntroCl4}, to the best of our knowledge, a formal study of the effect of the interaction of quantum states of light with such nanostructures is missing. 

In this work, we provide a framework to study the interaction between quantum states of light and a nanostructure. Our approach is completely general, although here we focus on an experimentally relevant situation: the scattering of two-photon states of light by a rotationally symmetric nanostructure. The symmetry of the problem allows us to focus on electromagnetic modes with well-defined total angular momentum $m=l +s$, where $l$ and $s$ represent the orbital and spin angular momentum, respectively. These types of quantum states are very robust to propagation \cite{Sciarrino15, ibrahim2013orbital, krenn2014communication} and can encode more information than for example polarization states by using several values of $l$ \cite{Oemrawsingh05, Franke08, Molina04, Molina07}. Interestingly, they can be manipulated in a controlled manner with rotationally symmetric nanostructures that allow for the conservation of the total angular momentum $m$ of the incoming light  \cite{Lasa20, Buse18}.

The theoretical framework used to describe the quantum scattering process is based on an input/output general formalism \cite{Tischler18, Barnett98, Buse18, vest2017anti}. We consider that the input and output states of the system are quantum states of light composed by two entangled photons, where the two photons have $m = 0$, and the information is encoded in their helicity $\Lambda$ (defined as the projection of the spin $s$ in the direction of propagation), which takes $\Lambda = +1$ or $\Lambda = -1$ values \cite{Buse18}. We consider a basis of four input two-photon modes which completely describe any input monochromatic two-photon states:
\begin{align}
&\ket{\psi_{\pm}^{i}(\omega_1, \omega_2)} = \frac{1}{2}\left\{\hat{a}_{i}^\dagger(\omega_1) \hat{a}_{i}^\dagger(\omega_2) \pm \hat{b}_{i}^\dagger(\omega_1)  \hat{b}_{i}^\dagger(\omega_2) \right\} \ket{0}, \label{eqStatesPsi}\\
&\ket{\chi_{\pm}^{i}(\omega_1, \omega_2)} = \frac{1}{2}\left\{\hat{a}_{i}^\dagger(\omega_1) \hat{b}_{i}^\dagger(\omega_2) \pm \hat{b}_{i}^\dagger(\omega_1) \hat{a}_{i}^\dagger(\omega_2)\right\}\ket{0}, \label{eqStatesChi}
\end{align}
where $\ket{0}$ is the vacuum state, and $\omega_1$ and $\omega_2$ are the frequencies of the two-photons. The basis of the two-photon output monochromatic modes also has four elements, $\ket{\psi_{\pm}^{o}(\omega_1, \omega_2)}$ and $\ket{\chi_{\pm}^{o}(\omega_1, \omega_2)}$ that follow Eqs. \eqref{eqStatesPsi} and \eqref{eqStatesChi}, respectively, but the input ``$i$'' labels are substituted by the output ``$o$'' labels. The modes of light are described by the input(output) $\hat{a}^\dagger_{i(o)}(\omega)$ and $\hat{b}^\dagger_{i(o)}(\omega)$ operators that indicate the creation of an input(output) photon with helicity $\Lambda = +1$ or $\Lambda = -1$, respectively. The $\hat{a}^\dagger_{i(o)}(\omega)$ and $\hat{b}^\dagger_{i(o)}(\omega)$ bosonic operators satisfy the canonical commutation relations \cite{CohenTannoudji_vol1} and are evaluated for a strictly monochromatic (angular) frequency $\omega$.

In experiments, the incident photon pairs are usually generated in a frequency superposition of states by a standard spontaneous parametric down-conversion (SPDC) process. This can result in a loss of purity after scattering off a nanostructure \cite{Buse18, schotland2016scattering}. We focus on the monochromatic mode $\ket{\psi_{+}^{i}(\omega_1, \omega_2)}$ and describe the incident two-photon state as a frequency superposition (for all possible input states see Supplemental Material \cite{supplementary}),
\begin{equation}
\begin{split}
    \ket{\Psi^{i}_{+}} = \iint d\omega_1 d\omega_2 \phi(\omega_1, \omega_2) \ket{\psi^{i}_{+}(\omega_1, \omega_2)}.
\end{split}
\label{eqPSI_IN}
\end{equation}
where $\phi(\omega_1, \omega_2)$ is the two-photon spectral function that we approximate as the product of two Gaussian functions centered both at the central frequency $\omega_\text{in}$ (or central wavelength $\lambda_\text{in} = 2 \pi c / \omega_\text{in}$) and variance $\sigma = 3$ THz (we choose a $\sigma$ value similar to the one used in recent experiments\cite{JJMiguel21, Tischler15}),
\begin{align}
    \nonumber
    &\phi(\omega_1, \omega_2) =\\
    &\frac{1}{\sigma \sqrt{\pi}} \exp\left(-\frac{(\omega_1 - \omega_\text{in})^2}{2 \sigma^2}\right) \exp\left(-\frac{(\omega_2 - \omega_\text{in})^2}{2 \sigma^2}\right).
    \label{eqphi}
\end{align}


\begin{figure}[t]
    \centering
    \includegraphics[width = 8.6 cm]{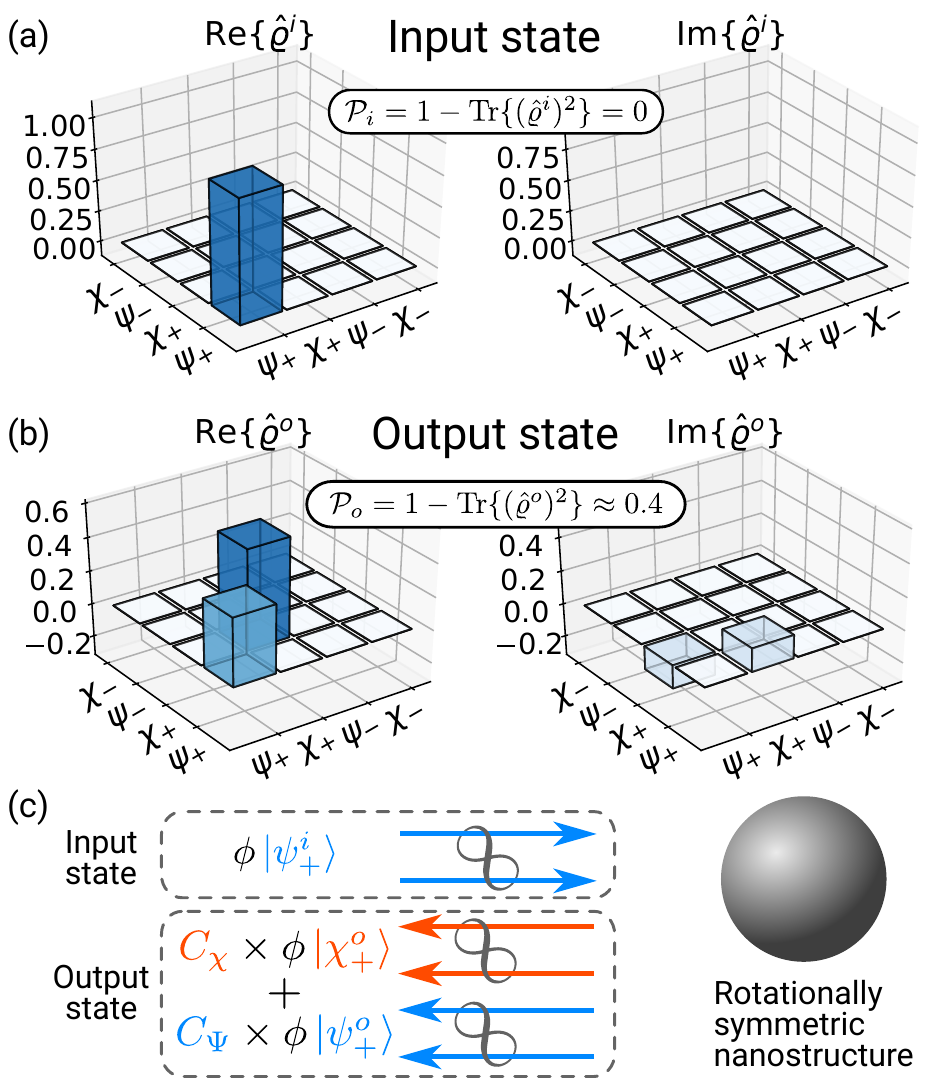}
    \caption{Scattering of two-photon quantum states by a nanostructure. (a) Real and imaginary components of $\hat{\varrho}^i$, the density matrix associated to the input state $\ket{\Psi^i_+}$. The two-photon spectral function is centered at $\omega_\text{in} = 17.5 \times 10^{14} $ rad/s and has a variance of $\sigma = 3$ THz. $\mathcal{P}_i = 0$ indicates that the input state is pure. (b) Real and imaginary components of $\hat{\varrho}^o$, the post-selected density matrix of the output state $\ket{\Psi^o_+}$ that results from the scattering of the incident input state $\ket{\Psi^i_+}$ in (a). For the calculation, we chose that the helicity splitting coefficients $\beta(\omega) = 0.2$ and $\alpha(\omega)$ a Lorentzian function with $\omega_L = 17.5 \times 10^{14}$ rad/s and $\gamma = 1 $ THz. $\mathcal{P}_o > 0$ indicates that the output state is not pure. (c) Scheme of the scattering process. The $\ket{\Psi^i_+}$ input state is scattered as a superposition of $\ket{\psi^o_{+}}$ and $\ket{\chi^o_{+}}$ with amplitudes given by $C_\psi \phi$ and $C_\chi \phi$, respectively.}
    \label{fig1}
\end{figure}

In order to approach a realistic experimental detection of quantum states, we consider that the scattered states are measured through a post-selection of two-photon states \footnote{The post-selection consists in ignoring the scattered states with less than two-photons.}. We also consider in the following that the detectors are ``blind'' to the frequency degree of freedom. Then, the input and output quantum states are best described with the experimentally-accessible density matrix, $\hat{\varrho}$, resulting after tracing out the frequency degree of freedom. Thus, the elements $\braket{\xi|\hat{\varrho}|\xi'}$ of the density matrix corresponds to the results of standard quantum state tomography measurements \cite{TomographyKwiat05, CohenTannoudji_vol1},
\begin{align}
    \nonumber
    &\braket{\xi|\hat{\varrho}^{i(o)}|\xi'} =\\
    &K \iint d\omega_1 d\omega_2 \braket{\xi(\omega_1, \omega_2)\ket{\Psi_+^{i(o)}} \bra{\Psi_+^{i(o)}} \xi'(\omega_1, \omega_2)},
    \label{eqRHO}
\end{align}
where $K$ is a normalization constant that ensures $\text{Tr}\{\hat{\varrho}^{i(o)}\} = 1$. $\ket{\xi(\omega_1, \omega_2)}$ and $\ket{\xi'(\omega_1, \omega_2)}$ can be any of the $\ket{\psi_\pm^{i(o)}(\omega_1, \omega_2)}$ and $\ket{\chi_\pm^{i(o)}(\omega_1, \omega_2)}$ states given in Eqs. \eqref{eqStatesPsi} and \eqref{eqStatesChi}, respectively. Figure \ref{fig1}a shows the density matrix $\hat{\varrho}^{i}$ of the incident state $\ket{\Psi^{i}_{+}}$ (Eq. \eqref{eqPSI_IN}) calculated using Eq. \eqref{eqRHO}. $\hat{\varrho}^{i}$ is characterized by a single non-zero element corresponding to $\braket{\psi_+|\hat{\varrho}^{i}|\psi_+}$, and there is no contribution from the other elements of the basis (Eqs. \eqref{eqStatesPsi} and \eqref{eqStatesChi}). The purity of the quantum state can then be quantified using $\mathcal{P}_{i(o)}=1-\text{Tr}\{(\hat{\varrho}^{i(o)})^2\}$, with $\mathcal{P}_{i(o)}=0$ indicating a pure state (and $\mathcal{P}_{i(o)}=1-1/N$ indicating a maximally mixed state, with $N$ the number of states considered). For example, the input density matrix $\hat{\varrho}^{i}$ in Fig.~\ref{fig1}a satisfies $\mathcal{P}_i=1-\text{Tr}\{(\hat{\varrho}^{i})^2\} = 0$, which confirms that $\hat{\varrho}^{i}$ is a pure state \cite{CohenTannoudji_vol1}.


We next apply the input/output formalism to analyse the loss of purity due to the scattering of $m = 0$ photons by a rotationally symmetric nanostructure. Rotationally symmetric structures conserve the total angular momentum of the incident light. However, the total angular momentum conservation does not imply the conservation of the vectorial degree of freedom of light, which is determined by the helicity $\Lambda$. Thus, the input electromagnetic modes with $m=0$ and $\Lambda = +1$ or $\Lambda = -1$ can only be scattered into two different output electromagnetic modes with $m = 0$ and $\Lambda = +1$ or $\Lambda = -1$. Further, we consider that photons can be lost or dissipated in the scattering process. This situation where two input electromagnetic modes are either lost or transformed into two other output electromagnetic modes is analogous to the situation of a lossy beam splitter \cite{Barnett98, Tischler18, vest2017anti}. Thus, we adapt the non-unitary transformation of lossy beam splitters to our system, resulting in an equation for the creation operators
\begin{align}
\nonumber
    \hat{a}_{o}^\dagger(\omega) &= \alpha^*(\omega) \hat{a}_{i}^\dagger(\omega) + \beta^*(\omega) \hat{b}_{i}^\dagger(\omega) + \hat{L}_{a}^\dagger(\omega), \\
    \hat{b}_{o}^\dagger(\omega) &= \alpha^*(\omega) \hat{b}_{i}^\dagger(\omega) + \beta^*(\omega) \hat{a}_{i}^\dagger(\omega) + \hat{L}_{b}^\dagger(\omega),
\label{eqTRANS}
\end{align}
where $\hat{L}_{a}^\dagger(\omega)$ and $\hat{L}_{b}^\dagger(\omega)$ are the Langevin operators accounting for the losses in the scattering process \cite{Barnett98, matloob1995electromagnetic, matloob1996electromagnetic}, and $\alpha(\omega)$ and $\beta(\omega)$ (conjugated in Eq. \eqref{eqTRANS}) are the helicity-splitting coefficients. Equation \eqref{eqTRANS} describes the scattering of quantum states of light, but the $\alpha(\omega)$ and $\beta(\omega)$ coefficients can be calculated from the classical response of the system obtained from Maxwell's equations. Experimentally, $\alpha(\omega)$ and $\beta(\omega)$ can be measured by using classical beams of light. Thus, Maxwell's equations determine how the $m = 0$ $\Lambda = \pm1$ electromagnetic modes get transformed, both in the classical and quantum regimes \cite{Birula94, schotland2016scattering}. 




From Eqs. \eqref{eqPSI_IN} and \eqref{eqTRANS} we obtain the expression of the output scattered state
\begin{align}
    \nonumber
    \ket{\Psi^o_{+}} &= \iint d\omega_1 d\omega_2 \phi(\omega_1, \omega_2) C_{\psi}(\omega_1, \omega_2)\ket{\psi^o_{+}(\omega_1, \omega_2)}\\
    &+\iint d\omega_1 d\omega_2 \phi(\omega_1, \omega_2) C_{\chi}(\omega_1, \omega_2)\ket{\chi^o_{+}(\omega_1, \omega_2)}.
\label{eqPSI_OUT}
\end{align}
$C_{\psi}(\omega_1, \omega_2)$ and $C_{\chi}(\omega_1, \omega_2)$ in Eq. \eqref{eqPSI_OUT} are frequency-dependent functions constructed from $\alpha(\omega)$ and $\beta(\omega)$ (the full expression can be found in the Supplemental Material \cite{supplementary}). This result shows that, in general, the output state is a superposition of two different entangled photon modes. The $C_{\psi}(\omega_1, \omega_2)$ and $C_{\chi}(\omega_1, \omega_2)$ coefficients may have a strong frequency dependence due to the interaction with the nanostructure, greatly affecting the output state. Using Eq.\eqref{eqPSI_OUT} we can calculate the density matrix $\hat{\varrho}^{o}$ (Eq. \eqref{eqRHO}). For example, Fig. \ref{fig1}b shows the output density matrix $\hat{\varrho}^{o}$, when we artificially set $\beta(\omega) = 0.2$ and $\alpha(\omega) = \omega_L \gamma / [2 (\omega_L^2 - \omega^2 + i \gamma \omega)]$ a Lorentzian function that mimics the resonant behavior of a nanostructure. The $\hat{\varrho}^{o}$ obtained for the output state represents a partially coherent superposition of $\ket{\psi^o_{+}(\omega_1, \omega_2)}$ and $\ket{\chi^o_{+}(\omega_1, \omega_2)}$ (as indicated by Eq. \eqref{eqPSI_OUT}). The purity of this state is $\mathcal{P}_o\approx 0.4 > 0$. This simple example demonstrates the process of purity loss of a quantum state interacting with a nanostructure.


We now identify the origin of this loss of purity. Let us consider that the input two-photon spectral function, $\phi(\omega_1, \omega_2)$ is quasi-monochromatic (\textit{i.e.,} its spectral variance $\sigma$ is significantly smaller than the central frequency of the pulse, $\omega_\text{in}$). Under this approximation, we can expand the $\phi(\omega_1, \omega_2) C_\psi(\omega_1, \omega_2)$ and $\phi(\omega_1, \omega_2) C_\chi(\omega_1, \omega_2)$ functions (Eq. \eqref{eqPSI_OUT}) to first order around the central frequency of the two-photon spectral function $\omega_\text{in}$:
\begin{widetext}
\begin{align}
    \phi(\omega_1, \omega_2) C_\psi(\omega_1, \omega_2) \approx A_\psi \exp\left[ -\frac{[\omega_1 - (\omega_\text{in} + \sigma^2 F_\Psi)]^2}{2\sigma^2}  + i \omega_1 \tau_\psi\right] \exp\left[ -\frac{[\omega_2 - (\omega_\text{in} + \sigma^2 F_\Psi)]^2}{2\sigma^2}  + i \omega_2 \tau_\psi\right],
    \label{eqPhiCpsi}
\end{align}
\begin{align}
    \phi(\omega_1, \omega_2) C_\chi(\omega_1, \omega_2) \approx A_\chi \exp\left[ -\frac{[\omega_1 - (\omega_\text{in} + \sigma^2 F_\chi)]^2 }{2\sigma^2} + i \omega_1 \tau_\chi\right] \exp\left[ -\frac{[\omega_2 - (\omega_\text{in} + \sigma^2 F_\chi)]^2}{2\sigma^2}  + i \omega_2 \tau_\chi\right],
    \label{eqPhiCchi}
\end{align}
\end{widetext}

\noindent
where $A_\Psi$, $F_\Psi$, $\tau_\Psi$, $A_\chi$, $F_\chi$, and $\tau_\chi$ are functions of $\alpha(\omega_\text{in})$, $\beta(\omega_\text{in})$, $(d\alpha/d\omega)|_{\omega_\text{in}}$, and $(d\beta/d\omega)|_{\omega_\text{in}}$ (see Supplemental Material \cite{supplementary}), \textit{i.e.} these parameters depend only on the classical response of the system evaluated at the central frequency of the incident pulse, $\omega_\text{in}$.


Eqs. \eqref{eqPhiCpsi} and \eqref{eqPhiCchi}, show that the output two-photon modes can still be described as two factorized Gaussian pulses, one for each photon. However, the amplitude ($A_\Psi$ and $A_\chi$), central frequency ($(\omega_\text{in} + \sigma^2 F_\Psi)$ and $(\omega_\text{in} + \sigma^2 F_\chi)$) and delay ($\tau_\psi$ and $\tau_\chi$) are, in general, different for $\phi(\omega_1, \omega_2) C_\psi(\omega_1, \omega_2)$ and $\phi(\omega_1, \omega_2) C_\chi(\omega_1, \omega_2)$. This implies that after the interaction with the nanostructure, the resulting quantum state is a superposition of two different quantum states with different time-frequency properties. In Fig. \ref{fig1}c we sketch the scattering process just described by these equations.



Using Eqs. \eqref{eqRHO}, and \eqref{eqPSI_OUT}-\eqref{eqPhiCchi} we obtain an analytical expression for the purity of the output state,
\begin{equation}
\begin{split}
    &\mathcal{P}_o= \\
    &\frac{2|A_\Psi|^2|A_\chi|^2e^{\sigma^2(F_\chi^2 + F_\Psi^2)}}{(|A_\Psi|^2 e^{2 \sigma^2 F_\Psi^2} + |A_\chi|^2 e^{2\sigma^2 F_\chi^2})^2} [e^{\sigma^2 \Delta F^2 } - e^{-\sigma^2 \Delta \tau^2}],
\end{split}
\label{eqImpurityApprox}
\end{equation}
with $\Delta F = F_\Psi - F_\chi$ and $\Delta \tau = \tau_\Psi - \tau_\chi$. This equation indicates that in a purely monochromatic situation ($\sigma \to 0$) the output state is pure ($\mathcal{P}_o = 0$), and it is a superposition of two different states with amplitudes $A_\Psi$ and $A_\chi$ (Eqs. \eqref{eqPSI_OUT}-\eqref{eqPhiCchi}). For non-monochromatic input states, if any of these amplitudes is zero the output state is again a pure state. This happens when $\beta(\omega_\text{in})  = 0$ (condition of helicity preservation), or $\alpha(\omega_\text{in}) = 0$ (condition of total conversion of helicity), or $\alpha(\omega_\text{in}) = \pm \beta(\omega_\text{in})$. The latter condition corresponds to the situation where the incident light only excites either magnetic or electric modes of the nanostructure (see Supplemental Material \cite{supplementary}) \cite{Zambrana13, olmos2020kerker, lasa2022correlations}. Additionally, the output state is also pure if $\Delta F = \Delta \tau = 0$. It can be shown that this case holds approximately when $\alpha(\omega)$ and $\beta(\omega)$ are almost constant in a spectral range given by $\sigma$. Then, a substantial loss of purity can be observed if the incident light excites both electric and magnetic resonances of the nanostructure, where $\alpha(\omega)$ and $\beta(\omega)$ change abruptly at the same time.  


\begin{figure}[t]
    \centering
    \includegraphics[width = 8.6 cm]{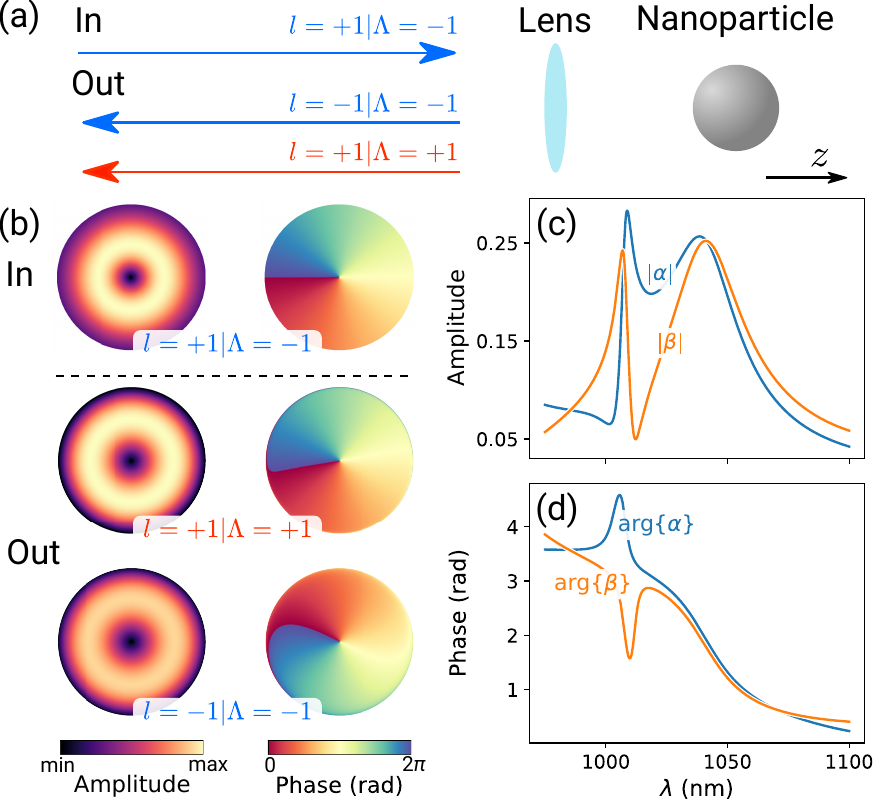}
    \caption{Classical response of a spherical nanoparticle illuminated by a beam with $m = 0$ and $\Lambda = -1$. (a) Scheme of the classical scattering. A Laguerre-Gauss beam with $m = 0$ and $\Lambda = -1$ ($l = +1$) propagates along the positive $z$-axis towards a $r = 250$ nm silicon \cite{Palik} spherical nanoparticle situated in vacuum. The incident beam is focused on the centre of the nanoparticle with a high numerical aperture lens ($NA = 0.9$) and focal length $f = 1$mm. The light backscattered by the nanoparticle is collimated with the same lens and separated into the two different helicities $\Lambda = +1$ and $\Lambda = -1$. (b) Spatial distribution of the amplitude (left column) and phase (right column) of the input and output electric fields for illumination wavelength $\lambda = 1000$ nm. The input and output fields are both plotted in the same plane of the aperture of the lens, situated before the input beam is focused. The input electric field (first row) is normalized to its maximum value. The output beam (second and third row) is decomposed into the two different contributions with different helicities, as indicated in the figure. The amplitude of both output contributions is normalized to the maximum of the output field with helicity $\Lambda = +1$. (c) Amplitude and (d) phase of the helicity-splitting coefficients $\alpha$ (blue line) and $\beta$ (orange line) that are obtained from the classical (scattering) calculation, as a function of wavelength $\lambda$.}
    \label{fig2}
\end{figure}

To illustrate a realistic scenario that can produce a loss of purity in the scattering process, we study the scattering of the quantum state $\ket{\Psi_+^i}$ (Eq. \eqref{eqPSI_IN}) by a silicon spherical nanoparticle of radius $r = 250$ nm in vacuum (the permittivity of silicon is obtained from \cite{Palik}). By applyng Mie's theory, we calculate the helicity-splitting coefficients $\alpha(\omega)$ and $\beta(\omega)$ under a $m = 0$, $\Lambda = -1$ illumination mode \cite{Zambrana12, Gabriel08, supplementary}. We use a set of parameters close to those in experimental configurations: we consider an incident Laguerre-Gauss mode with spatial width $w_0 = 0.5$ mm and a high numerical aperture lens ($NA = 0.9$) with focal length $f = 1$ mm to focus the incident light at the centre of the nanoparticle (see scheme in Fig. \ref{fig2}a). Figure \ref{fig2}b (upper panel) shows the spatial distribution of the phase and intensity of the input beam for an incident monochromatic illumination with $\lambda = 1000$ nm.  The lens used to focus the incident beam also collects the backscattered light, which is then separated into two contributions, one with $\Lambda = +1$ and another $\Lambda = -1$ (scheme in Fig. \ref{fig2}a). Note that $\Lambda = +1$ and $m = 0$ corresponds to $l = -1$ for the incident beam and $l = +1$ for the backscattered beam due to the opposite propagation direction. Figure \ref{fig2}b (lower panel) shows the spatial field distribution of the two helicity contributions of the output beam after being collimated by the lens. To obtain the values of $\alpha(\omega)$ and $\beta(\omega)$ we project the spatial field distribution of the output field onto a Laguerre-Gauss mode with $\Lambda = -1$ ($l = -1$) and $\Lambda = +1$ ($l = +1$), respectively. 

We show in Fig. \ref{fig2}c-d the amplitude and phase spectra of $\alpha(\omega)$ (blue line) and $\beta(\omega)$ (orange line), obtained by performing the projection of the output fields for different wavelengths $\lambda \in [975, 1150]$ nm. Figure \ref{fig2}c shows the amplitudes $|\alpha(\omega)|$ and $|\beta(\omega)|$ with two clear peaks: a relatively broad maxima centred at $\lambda \approx 1040$ nm and a sharp peak centred at $\lambda \approx 1000$ nm. In the proximity of these two peaks the phase of the coefficients ($\text{arg}\{\alpha(\omega)\}$ and $\text{arg}\{\beta(\omega)\}$) varies rapidly, particularly near the sharp peak (Fig. \ref{fig2}d). This behavior is due to the excitation of two optical resonances of the silicon nanoparticle, a magnetic octopole centred at $\lambda \approx 1007$ nm and an electric quadrupole centred at $\lambda \approx 1040$ nm (see Supplemental Material \cite{supplementary}).

\begin{figure}[t!]
    \centering
    \includegraphics[width = 8.6 cm]{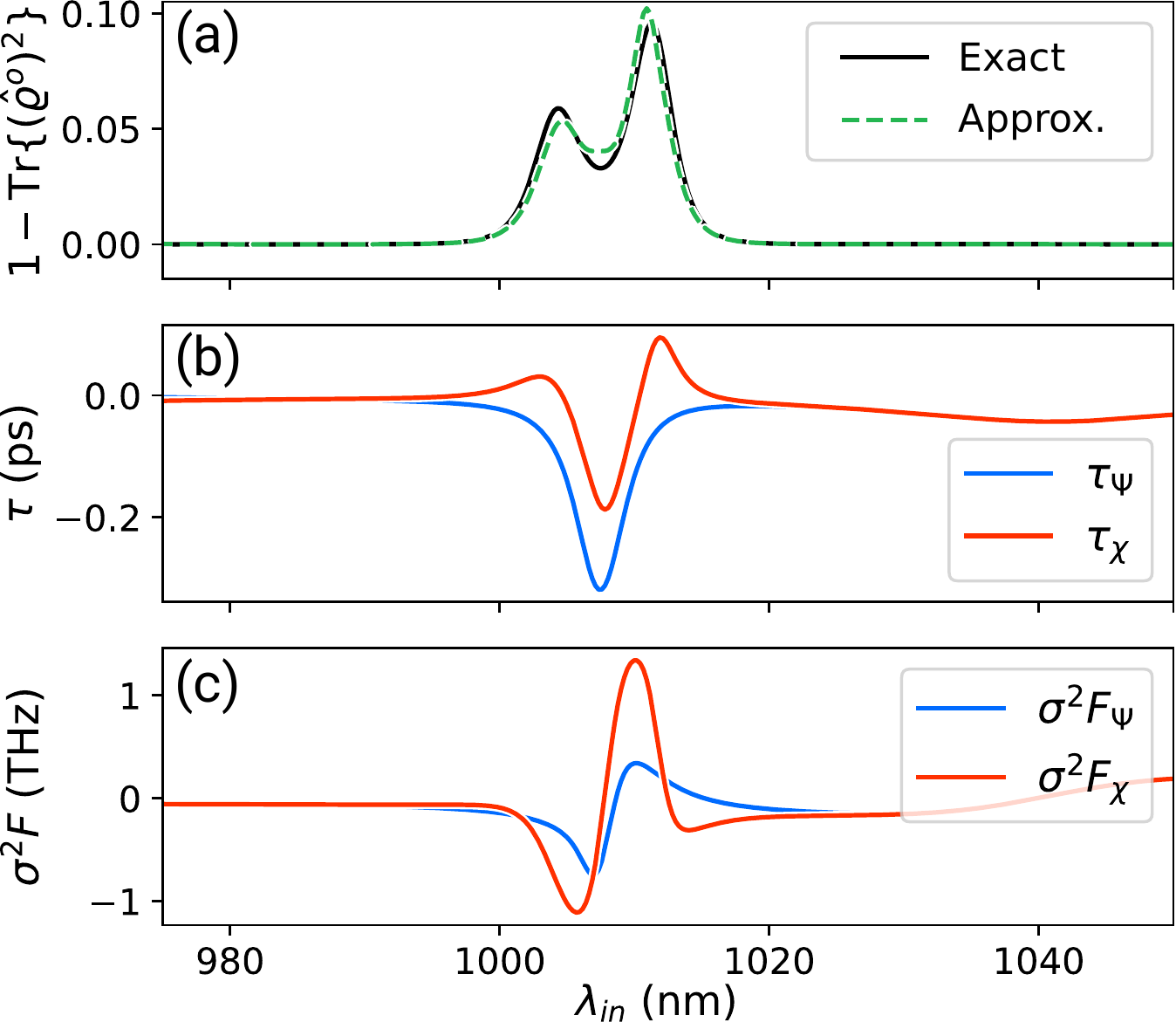}
    \caption{Analysis of the purity loss for an incident $\ket{\Psi_+^i}$ state interacting with a $r = 250$ nm silicon \cite{Palik} spherical nanoparticle in vacuum. (a) Spectrum of the purity loss $1 - \text{Tr}\{(\hat{\varrho}^{o})^2\}$. The solid black line corresponds to the purity loss of the numerically calculated density matrix (Eq. \eqref{eqRHO}) of the output state (Eq. \eqref{eqPSI_OUT}) and the dashed green line to the evaluation of Eq. \eqref{eqImpurityApprox}. (b) Spectrum of the time delay of the two two-photon modes of the total output state (Eq. \eqref{eqPSI_OUT}). The solid blue line corresponds to $\tau_\Psi$ and the solid red line to $\tau_\chi$. (c) Spectrum of the frequency-shift of the two two-photon modes of the total output state. The solid blue line corresponds to $\sigma^2 F_\Psi$ and the solid red line to $\sigma^2 F_\chi$.
    }
    \label{fig3}
\end{figure}

With the results of $\alpha(\omega)$ and $\beta(\omega)$ shown in Fig. \ref{fig2}c-d we can now calculate the post-selected output density matrix $\hat{\varrho}^o$ (using Eqs. \eqref{eqRHO}-\eqref{eqPSI_OUT}) when the system is illuminated by the entangled input state $\ket{\Psi^{i}_{+}}$. We consider again a Gaussian frequency mode with a width $\sigma = 3$ THz. Figure \ref{fig3}a (solid black line) shows the purity of the output state $\mathcal{P}_o$ as a function of the central wavelength of the pulse $\lambda_\text{in} = 2 \pi c / \omega_\text{in}$ ($c$ being the speed of light in vacuum). We find two clear peaks situated at $\lambda_\text{in} \approx 1003$ nm and $\lambda_\text{in} = 1012$ nm where the purity loss becomes significant ($\mathcal{P}_o \sim 0.1$).


To analyze this result we consider that the output state is the superposition of two time-delayed and frequency-shifted states \cite{rangarajan2009optimizing} (see the sketch in Fig. \ref{fig1}c and Eqs. \eqref{eqPSI_OUT}-\eqref{eqPhiCchi}). In Fig. \ref{fig3}b we show the spectral dependence of the time delays $\tau_\Psi$ (solid blue line) and $\tau_\chi$ (solid red line), and in Fig. \ref{fig3}c we show the corresponding values of the frequency shifts $\sigma^2 F_\Psi$ and $\sigma^2 F_\chi$. In Fig. \ref{fig3}a (green dashed line) we show the result of evaluating Eq. \eqref{eqImpurityApprox} with the values of $\tau_\Psi$, $\tau_\chi$, $\sigma^2 F_\Psi$, and $\sigma^2 F_\chi$ shown in Figs. \ref{fig3}b and \ref{fig3}c. We observe that there is a very good agreement between the results obtained with the approximated expression, and the exact expression of the loss of purity. As expected from Eq. \eqref{eqImpurityApprox} the loss of purity becomes maximized where the differences between $\tau_\Psi$ and $\tau_\chi$ and between $\sigma^2 F_\Psi$ and $\sigma^2 F_\chi$ are larger, as shown in the region $1000$ nm $\lesssim\lambda_\text{in}\lesssim 1020$ nm of Figs. \ref{fig3}a-c. Further, we observe by comparing Figs. \ref{fig2}c-d with Figs. \ref{fig3}b-c that the larger time delays and frequency shifts are obtained in the region where the spectral variations of $|\alpha(\omega)| - |\beta(\omega)|$ and $\text{arg}\{\alpha(\omega)\} - \text{arg}\{\beta(\omega)\}$ are pronounced within the spectral width of the incident pulse. These abrupt spectral variations are due to the simultaneous excitation of both magnetic octopole and electric quadrupole resonances of the nanoparticle, clearly pointing out towards the influence of the optical resonances of the nanostrucutre in the loss of purity.

In summary, in this work we show that the interaction of entangled photons with nanostructures can produce a loss of purity of the incident state, even when the nanostructure is smaller than the wavelength of the incident field. In particular, we provide a framework to treat the quantum scattering of a two-photon state with modes of well-defined angular momentum and helicity interacting with a rotationally symmetric nanostructure, where we can connect the loss of purity with the excitation of different optical resonances of the nanoparticle. We apply our framework to analyze the loss of purity in a completely realistic scenario, a silicon spherical nanoparticle illuminated by a focused two-photon state generated by a SPDC process. Further, we developed a simple physical picture that explains the loss of quantum purity as a consequence of a time delay and a frequency shift between the two-photon modes of the output state. Our framework and analysis can be extended to other systems in order to find the loss of purity of incident quantum states interacting with a nanostructure; either to avoid the conditions that result in a loss of purity for quantum information applications or to exploit the loss of purity for characterizing the properties of the nanostructure.

\begin{acknowledgments}
\'{A}.N., R.E., and J.A. acknowledge financial support from the Spanish Ministry of Science and Innovation and the Spanish goverment agency of research MCIN/AEI/10.13039/501100011033 through Project Ref. No. PID2019-107432GB-100. Á.N., R.E., C.M.-E. and J.A. acknowledge financial support from the Department of Education, Research and Universities of the Basque Government through Project Ref. No. IT-1526-22. This work has been also supported by the CSIC Interdisciplinary Thematic Platform (PTI+) on Quantum Technologies (PTI-QTEP+).
\end{acknowledgments}

\nocite{*}

\bibliography{bibliography}

\end{document}


\maketitle

\noindent\makebox[\linewidth]{\rule{\textwidth}{0.4pt}}
\tableofcontents
\noindent\makebox[\linewidth]{\rule{\textwidth}{0.4pt}}
\newpage

\section{Details of the classical scattering calculations}
\label{secS1}

The interaction between quantum states of light and a spherical nanoparticle is described by Eq. (6) in the main text, and depends on the $\alpha(\omega)$ and $\beta(\omega)$ helicity-splitting coefficients. This section describes all the steps required to obtain these two coefficients for the system studied in the main text. 

$\alpha(\omega)$ and $\beta(\omega)$ can be obtained from the analysis of the fields scattered by the nanostructure when it is illuminated by classical light characterized by the same spatial dependence of the electromagnetic fields as the quantum states considered. We can use the classical response as an input to later describe the quantum behavior because Maxwell's equations determine how the electromagnetic modes of classical and quantized states of light evolve \cite{Birula94}.

The system that we study (see Fig. 2a of the main text) consists on a spherical nanoparticle that scatters a highly focused Laguerre-Gauss beam characterized by total angular momentum $m=0$ and spin $s = -1$ (and thus, orbital angular momentum $l = m - s = +1$). We note that the values of $\alpha(\omega)$ and $\beta(\omega)$ are independent of the chosen $s$ for $m = 0$ incident beams due to the mirror symmetry of the nanostrucutre. This input beam propagates along the positive $z$-axis and is focused by an aplanatic lens of a high numerical aperture $NA = 0.9$ placed at the focal length $f = 1$ mm from the center of the nanoparticle. The backscattered field (\textit{i.e.}, the scattering of the nanoparticle in the direction of negative $z$) is collected and collimated by the same lens. The collimated field is separated into two-helicity contributions. This separation can be achieved experimentally with a polarized beam splitter because the backscattered beam has a planar wavefront after the collimation, so that there is a direct relationship between the helicity $\Lambda$ and the polarization of the field \cite{tischler2014experimental}: the backscattered beams with left or right circular polarizations have spin $s = +1$ ($\Lambda = -1$) or $s = -1$ ($\Lambda = +1$), respectively ($\Lambda$ is defined as the spin projected in the direction of propagation). 

To obtain the $\alpha(\omega)$ and $\beta(\omega)$ helicity-splitting coefficients, we describe the field profile of the $m = 0$, $s = -1$ Laguerre-Gauss mode to be focused by the lens following the standard form \cite{siegman1986lasers}:

\begin{equation}
    \bm{E}_\text{LG}^{(\Lambda)}(\rho, \varphi_\text{p}) = \sqrt{\frac{2 q!}{\pi (q + |l|)!}} \frac{1}{w_0} \left( \frac{\rho \sqrt{2}}{w_0} \right)^{|l|} \exp\left( \frac{-\rho^2}{w_0^2} \right) L_q^l\left( \frac{2 \rho^2}{w_0^2} \right) \exp\left( i l \varphi_\text{p} \right) \hat{\bm{u}}_s, 
    \label{eqLGin}
\end{equation}
where $\rho$ and $\varphi_\text{p}$ are the radial and polar angle coordinates, respectively, at the aperture of the lens (with $\rho = 0$ being the center of the lens), $q$ is the number of radial nodes of the incident Laguerre-Gauss beam (for this work $q = 0$), $w_0 = 0.5$ mm is the width of the beam, $L_q^l$ are the generalized Laguerre polynomials \cite{GenLag}, and $\hat{\bm{u}}_s$ are the spin unitary vectors, which in Cartesian coordinates are written as $\hat{\bm{u}}_{+1} = (1_x, i_y, 0_z) / \sqrt{2}$, $\hat{\bm{u}}_{-1} = (-1_x, i_y, 0_z) / \sqrt{2}$, and $\hat{\bm{u}}_{+1} = (0_x, 0_y, 1_z)$. The field in Eq. \eqref{eqLGin} is normalized such that $\int_0^\infty d\rho \int_0^{2\pi} dp |E_{LG}^{(\Lambda)}|^2 = 1$ V$^2/$m$^2$.

\begin{figure}[t]
    \centering
    \includegraphics[width=8cm]{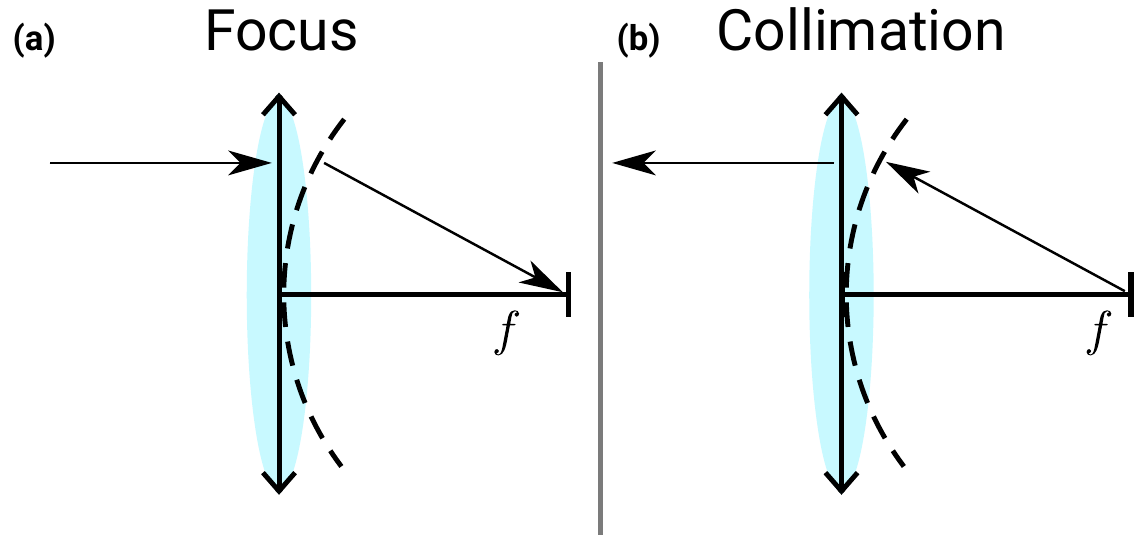}
    \caption{Scheme of the (a) focusing and (b) collimation process described by the aplanatic lens model. In the focusing process the field incident on the aperture of the lens (vertical arrow in the shaded blue area) is mapped into a spherical surface of radius $f$ (dashed line) and then rotated such that it propagates towards the focus (\textit{i.e.}, the center of this spherical surface). The collimation corresponds to the inverse process, so that the field is evaluated in the same spherical surface, rotated, and then mapped into the aperture of the lens.}
    \label{figS1}
\end{figure}

The beam given by Eq. \eqref{eqLGin} is focused by the lens. We follow Ref. \cite{Zambrana12} and write the focused field as an expansion of multipoles $\bm{A}_{1, n}^{(\Lambda)}$ of different order $n$, which facilitates the calculation of the scattered fields using Mie's theory. The index ``$1$'' in $\bm{A}_{1, n}^{(\Lambda)}$ indicates that this basis element is proportional to the spherical Bessel function of the first kind (see chapter 4 of Ref. \cite{BohrenHuffman}). The expansion considered here uses a basis such that each multipole, $\bm{A}_{1, n}^{(\Lambda)}$, has a well-defined helicity \cite{Zambrana12, Nora12, Rose, NovHecht}. The focused electric field is:

\begin{equation}
    \bm{E}_\text{foc}(r, \varphi_\text{s}, \theta, \omega) = \sum_{n = 0}^{\infty} \sqrt{2} C_n(\omega) \bm{A}_{1, n}^{(\Lambda)}(r, \varphi_\text{s}, \theta, \omega),
    \label{eqLGfoc}
\end{equation}
where $\omega$ is the angular frequency of the light, and $r$, $\varphi_\text{s}$, and $\theta$ are the spherical coordinates (radial, polar, and azimuthal, respectively) with origin at the center of the nanoparticle. The coefficients $C_n$ are given by (see Ref. \cite{Zambrana12}),

\begin{equation}
    C_n = i^n \sqrt{2n + 1} \times k \sqrt{2\pi} \int_0^{\theta_\text{max}} \sin(\theta) d\theta -i (\bm{E}_\text{LG}^{(\Lambda)}(f  \sin(\theta), 0) \cdot \hat{\bm{u}}_s) \sqrt{\cos(\theta)} f e^{-i k f},
    \label{eqCn}
\end{equation}
where $\theta_\text{max}$ is the maximal half-angle of the lens, and $k = 2 \pi /\lambda$, $\lambda$ being the wavelength of light in vacuum. These coefficients are derived in Ref. \cite{Zambrana12} using the aplanatic lens model \cite{NovHecht}, schematized in Fig. \ref{figS1}a. The modeling of the focusing process used to obtain Eq. \eqref{eqCn} can be separated into three steps. First, we map the incident electric field in the aperture of the lens onto an reference surface with coordinates $r = f$, $\varphi_\text{s} \in [0, 2 \pi]$, and $\theta \in [\pi - \theta_{max}, \pi]$ (\textit{i.e.} a spherical cap situated at the focal distance, $f$, from the nanoparticle). Second, the mapped (vectorial) electric field is rotated such that from each point of the  reference surface emerges a plane wave that propagates towards the focal point. Third, we obtain the field in the focal point as the sum of all these plane waves.

We next obtain the fields scattered by the nanoparticle using the expression \cite{Zambrana12}:

\begin{equation}
    \bm{E}_\text{SC}(r, \varphi_\text{s}, \theta, \omega) = \sum_{n = 0}^{\infty} C_n(\omega) \left( V_n(\omega) \bm{A}_{3, n}^{(\Lambda)}(r, \varphi_\text{s}, \theta, \omega) + W_n(\omega) \bm{A}_{3, n}^{(-\Lambda)}(r, \varphi_\text{s}, \theta, \omega) \right),
    \label{eqESC}
\end{equation}
where the multipoles $\bm{A}_{3, n}^{(\Lambda)}$ are in this case proportional to the spherical Bessel functions of the third kind, and

\begin{equation}
    V_n(\omega) = -\frac{a_n(\omega) + b_n(\omega)}{\sqrt{2}},
    \label{eqVn}
\end{equation}
\begin{equation}
    W_n = \frac{a_n(\omega) - b_n(\omega)}{\sqrt{2}}.
    \label{eqWn}
\end{equation}
$a_n(\omega)$ and $b_n(\omega)$ are the standard scattering coefficients in Mie's theory to describe the contributions of the electric and magnetic modes, respectively. The expressions of $a_n(\omega)$ and $b_n(\omega)$ can be found e.g. in Eq. (4.53) of Ref. \cite{BohrenHuffman}.


The backscattered field given in Eq. \eqref{eqESC} is collimated through the same lens that focuses the incident beam. To model the effect of this collimation using the aplanatic lens model, we follow the inverse process of the focusing described above (Fig. \ref{figS1}b). The backscattered field is evaluated at the spherical reference surface (dashed line in Fig. \ref{figS1}b). We then perform the inverse rotation than for the focusing process so that all the propagation vectors of the scattered field become aligned and perpendicular to the aperture of the lens. Finally, we map the rotated field of the reference aperture into the surface of the lens. This collimation process is mathematically equivalent to the following transformation:

\begin{equation}
    \bm{E}_\text{col} (\rho, \varphi_\text{p}, \omega) = \hat{R}(f, \varphi_\text{s}, \theta) \cdot \bm{E}_{SC}(f, \varphi_\text{s}, \theta, \omega) \cos(\theta)^{-1},
    \label{eqCol}
\end{equation}
where $\bm{E}_\text{col}$ is the collimated field, $\rho$ and $\varphi_\text{p}$ are the polar coordiantes in the aperture of the lens (defined above), and $\hat{R}(f, \varphi_\text{s}, \theta)$ is the position-dependent Euler rotation matrix:

\begin{equation}
    \hat{R}(f, \varphi_\text{s}, \theta) = \begin{pmatrix}
    \sin(\varphi_\text{s}) & - \cos(\varphi_\text{s}) & 0\\
    \cos(\varphi_\text{s}) & \sin(\varphi_\text{s}) & 0\\
    0 & 0 & 1
    \end{pmatrix} \cdot \begin{pmatrix}
    1 & 0 & 0\\
    0 & \cos(\theta) & - \sin(\theta)\\
    0 & \sin(\theta) & \cos(\theta)
    \end{pmatrix} \cdot \begin{pmatrix}
    \sin(\varphi_\text{s}) & \cos(\varphi_\text{s}) & 0\\
    - \cos(\varphi_\text{s}) & \sin(\varphi_\text{s}) & 0\\
    0 & 0 & 1
    \end{pmatrix}.
\end{equation}
The $\cos(\theta)^{-1}$ factor in Eq. \eqref{eqCol} accounts for the differences between the differential area at the reference spherical surface, $dA_S$, and the differential area at the aperture of the lens, $dA_L$ ($dA_S = dA_L / \cos(\theta)$, see chapter 3 of Ref. \cite{NovHecht}).

Finally, we obtain the helicity-splitting coefficients $\alpha(\omega)$ and $\beta(\omega)$ by projecting $\bm{E}_\text{col} (\rho, p, \omega)$ into the same input electromagnetic modes described in Eq. \eqref{eqLGin},

\begin{equation}
    \alpha(\omega) = \int_0^{2\pi} d p \int_0^D \rho d\rho [\bm{E}_\text{LG}^{(\Lambda)}(\rho, p)]^* \cdot \bm{E}_\text{col}(\rho, p, \omega),
    \label{eqalpha}
\end{equation}
and
\begin{equation}
    \beta(\omega) = \int_0^{2\pi} d p \int_0^D \rho d\rho [\bm{E}_\text{LG}^{(-\Lambda)}(\rho, p)]^* \cdot \bm{E}_\text{col} (\rho, p, \omega),
    \label{eqbeta}
\end{equation}
where $D = 2f \tan[\arcsin(NA)]$ is the lens diameter.

\subsection{Analysis of the different multipolar contributions}
\begin{figure}[t]
    \centering
    \includegraphics[width=8.6 cm]{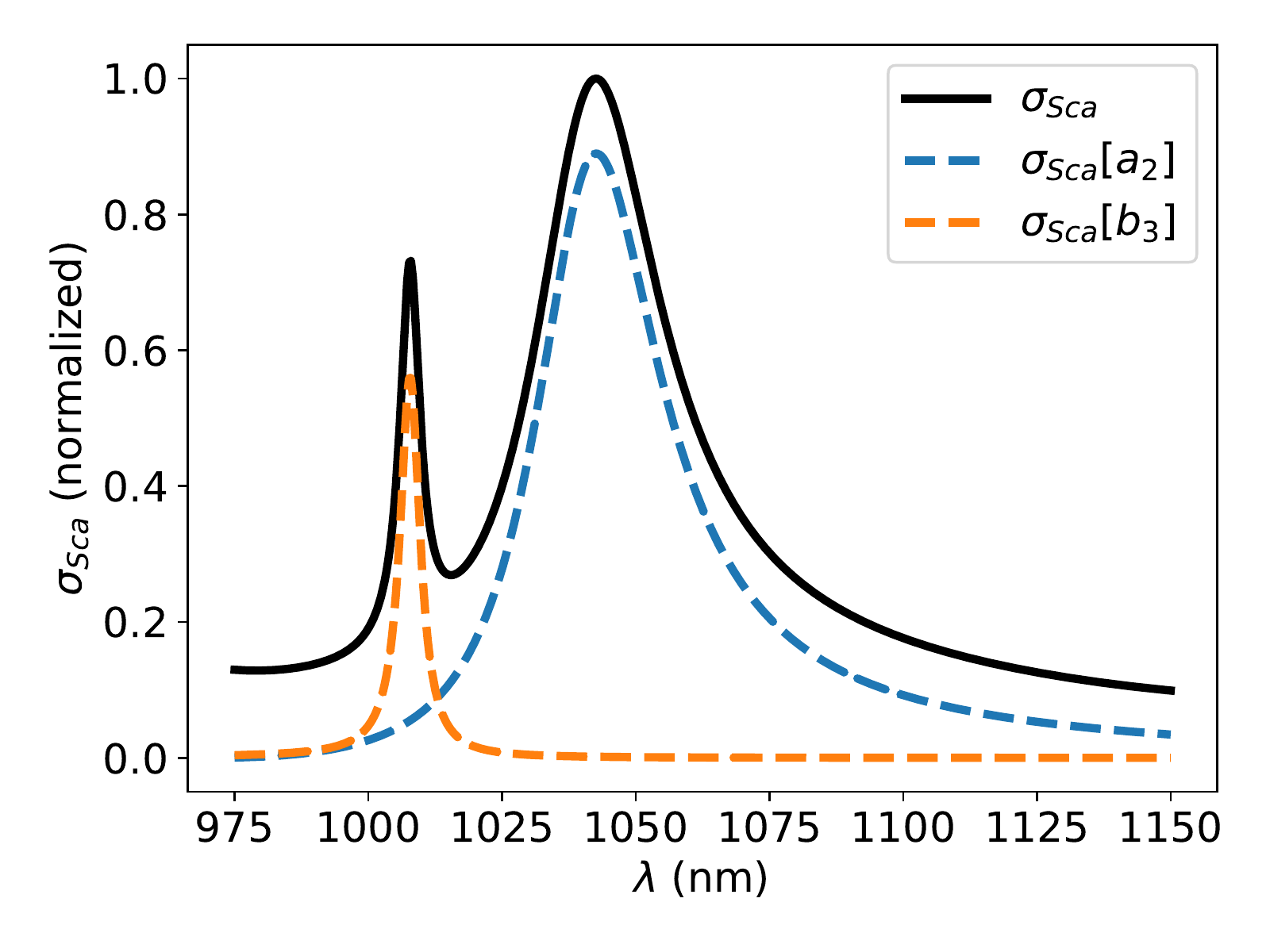}
    \caption{Different contributions to the cross section of the silicon \cite{Palik} spherical nanoparticle of radius $r = 250$ nm that is studied in the main text. The nanoparticle is illuminated by a Laguerre-Gauss beam with $l = 1$, $s = -1$, and width $w_0 = 0.5$mm that is focused by a lens of $NA = 0.9$ and $f = 1$mm at the center of the spherical nanoparticle. The solid black line corresponds to the total scattering cross section of the system $\sigma_{Sca}$ and the dashed lines to the two main contributions to $\sigma_{Sca}$; due to the electric quadrupolar ($\sigma_{Sca}[a_2]$, dashed blue line) and the magnetic octopolar ($\sigma_{Sca}[b_3]$, dashed orange line) modes of the nanoparticle.}
    \label{figS2}
\end{figure}

In this section, we analyze the nature of the peaks of $|\alpha(\omega)|$ and $|\beta(\omega)|$ in Fig. 2c of the main text. To that end, we show in Fig. \ref{figS2} (black line) the classical scattering cross-section spectrum $\sigma_\text{Sca}$ of the $r = 250$ nm silicon spherical nanoparticle studied in the main text (all other parameters can be found in the caption of Fig. \ref{figS2}). $\sigma_{Sca}$ has two clear peaks, a broad one centered at $\lambda \approx 1040$ nm and a narrow peak centered at $\lambda \approx 1010$ nm. The spectral position of the maximum of these two peaks is very close to the maximum of the two peaks of the $|\alpha(\omega)|$ and $|\beta(\omega)|$ helicity-splitting coefficients (shown in Fig. 2c of the main text). This indicates that the peaks of the helicity-splitting coefficients are due to the excitation of the resonances of the nanoparticles. To analyze which resonances are excited at this spectral positions we write $\sigma_{Sca}$ as a sum of the contributions of each multipole \cite{Zambrana12}:

\begin{equation}
    \sigma_{Sca} = \frac{2\pi}{k^2} \sum_{n = 0}^\infty |C_n(\omega)|^2 \left(|a_n(\omega)|^2 + |b_n(\omega)|^2\right) ,
    \label{eqsSca}
\end{equation}
where $C_n$ is given by Eq. \eqref{eqCn} and $a_n$ and $b_n$ (introduced in Eqs. \eqref{eqVn} and \eqref{eqWn}) are the scattering coefficients of the electric and magnetic multipoles, respectively.

In Fig. \ref{figS2}, we plot the two contributions given by the scattering coefficients $a_2$ (dashed blue line) and $b_3$ (dashed orange line) in Eq. \eqref{eqsSca}. The broad peak centered at $\lambda \approx 1040$ nm is due to the term proportional to $|a_2|^2$ in Eq. \eqref{eqsSca}, which indicates that this peak results from the excitation of an electric quadrupolar resonance in the nanoparticle \cite{BohrenHuffman}. The term proportional to $|b_3|^2$ is behind the narrow peak centered at $\lambda \approx 1010$ nm and corresponds to the excitation of an octopolar magnetic resonance in the nanoparticle \cite{BohrenHuffman}.

\section{Details of the quantum scattering calculations}
\label{secS2}
We derive in this section Eq. (7) of the main text, which gives the output state, $\ket{\Psi_+^o}$, scattered by a spherical nanoparticle for an incident two-photon state,

\begin{equation}
\begin{split}
    \ket{\Psi^{i}_{+}} = \iint d\omega_1 d\omega_2 \phi(\omega_1, \omega_2) \ket{\psi^{i}_{+}(\omega_1, \omega_2)},
\end{split}
\label{eqPSI_IN}
\end{equation}
where $\ket{\psi^{i}_{+}(\omega_1, \omega_2)}$ is given by Eq. (1) of the main text, and $\phi(\omega_1, \omega_2)$ is the two--photon spectral function. We consider that $\phi(\omega_1, \omega_2)$ is a product of two Gaussian pulses centered at frequency $\omega_\text{in}$ and with variance $\sigma$:

\begin{equation}
    \phi(\omega_1, \omega_2) = \frac{1}{\sigma \sqrt{\pi}} \exp\left(-\frac{(\omega_1 - \omega_\text{in})^2}{2 \sigma^2}\right) \exp\left(-\frac{(\omega_2 - \omega_\text{in})^2}{2 \sigma^2}\right).
    \label{eqphi}
\end{equation}
Note that this expression of $\phi(\omega_1, \omega_2)$ implies that the two photons are indistinguishable, since $\phi(\omega_1, \omega_2) = \phi(\omega_2, \omega_1)$.

To obtain the output state, we project the input state, $\ket{\Psi_+^i}$, on all the two-photon states of the output basis:
\begin{align}
    \nonumber
    \ket{\Psi^{o}_{+}} = 
    \left[ \ket{\psi_+^o(\omega_3, \omega_4)}\bra{\psi_+^o(\omega_3, \omega_4)} + \ket{\psi_-^o(\omega_3, \omega_4)}\bra{\psi_-^o(\omega_3, \omega_4)} + \right. \\
    \left. \ket{\chi_+^o(\omega_3, \omega_4)}\bra{\chi_+^o(\omega_3, \omega_4)} + \ket{\chi_-^o(\omega_3, \omega_4)}\bra{\chi_-^o(\omega_3, \omega_4)} \right] \iint d\omega_1 d\omega_2 \phi(\omega_1, \omega_2)  \ket{\psi^{i}_{+}(\omega_1, \omega_2)}.
    \label{eqPSI_O_0}
\end{align}

Next, we consider the relationship between the input and output modes \cite{Barnett98, Tischler18}:

\begin{align}
\nonumber
    \hat{a}_{o}^\dagger(\omega) &= \alpha^*(\omega) \hat{a}_{i}^\dagger(\omega) + \beta^*(\omega) \hat{b}_{i}^\dagger(\omega) + \hat{L}_{a}^\dagger(\omega), \\
    \hat{b}_{o}^\dagger(\omega) &= \alpha^*(\omega) \hat{b}_{i}^\dagger(\omega) + \beta^*(\omega) \hat{a}_{i}^\dagger(\omega) + \hat{L}_{b}^\dagger(\omega),
\label{eqTRANS}
\end{align}
where, $\alpha(\omega)$ and $\beta(\omega)$ are the helicity--splitting coefficients and the operators $\hat{L}_{a}^\dagger(\omega)$ and $\hat{L}_{b}^\dagger(\omega)$ account for the losses of the system.

To evaluate Eq. \eqref{eqPSI_O_0}, we substitute the expressions of the $\hat{a}_{o}^\dagger(\omega)$ and $\hat{b}_{o}^\dagger(\omega)$ operators in Eq. \eqref{eqTRANS} into the expressions of $\ket{\psi_+^o(\omega_1, \omega_2)}$, $\ket{\psi_-^o(\omega_1, \omega_2)}$, $\ket{\chi_+^o(\omega_1, \omega_2)}$, and $\ket{\chi_-^o(\omega_1, \omega_2)}$ (given by Eqs. (1) and (2) of the main text), \textit{i.e.}, we express the elements of the output basis in terms of the input operators. After some algebraic manipulation, Eq. \eqref{eqPSI_O_0} becomes:

\begin{align}
    \ket{\Psi^o_{+}} &= \iint d\omega_1 d\omega_2 \phi(\omega_1, \omega_2) \left[C_{\psi}(\omega_1, \omega_2)\ket{\psi^o_{+}(\omega_1, \omega_2)} + C_{\chi}(\omega_1, \omega_2)\ket{\chi^o_{+}(\omega_1, \omega_2)}\right],
\label{eqPSI_OUT}
\end{align}
with $C_{\psi}(\omega_1, \omega_2)$ and $C_{\chi}(\omega_1, \omega_2)$ defined as:

\begin{align}
    \label{eqCPsi0}
    C_\psi(\omega_1, \omega_2)  = {\alpha}(\omega_1){\alpha}(\omega_2) + {\beta}(\omega_1){\beta}(\omega_2),\\
    C_\chi(\omega_1, \omega_2) = {\alpha}(\omega_1){\beta}(\omega_2) + {\beta}(\omega_1){\alpha}(\omega_2),
    \label{eqCChi0}
\end{align}
which concludes the derivation of Eq. (7) of the main text.

\subsection{Quasi--monochromatic approximation}
In this section, we derive the quasi-monochromatic expressions of $C_{\psi}(\omega_1, \omega_2)$ and $C_{\chi}(\omega_1, \omega_2)$ given in Eqs. (8) and (9) of the main text. To that end, we first expand the $\alpha(\omega)$ and $\beta(\omega)$ helicity--splitting coefficients given by Eqs. \eqref{eqalpha} and \eqref{eqbeta}) to first order (around the central frequency of the pulses, $\omega_{in}$,

\begin{equation}
    \alpha(\omega) \approx A \left( 1 + \frac{A'}{A} \Delta\omega \right),
    \label{eqAexp}
\end{equation}
\begin{equation}
    \beta(\omega) \approx B \left( 1 + \frac{B'}{B} \Delta\omega \right),
    \label{eqBexp}
\end{equation}
with $A = \alpha(\omega_{in})$, $B = \beta(\omega_{in})$, $A' = d\alpha(\omega)/d\omega|_{\omega_{in}}$, $B' = d\beta(\omega)/d\omega|_{\omega_{in}}$, and $\Delta\omega = \omega - \omega_{in}$. This expansion is justified when the spectral changes of $\alpha(\omega)$ and $\beta(\omega)$ are small in the spectral range set by the variance of the incident pulse $\sigma$ \cite{JJMiguel21}.

Using Eqs. \eqref{eqAexp} and \eqref{eqBexp} we can write Eq. \eqref{eqCPsi0} and \eqref{eqCChi0} as:

\begin{equation}
    C_\psi(\omega_1, \omega_2) \approx (A^2 + B^2) [1 + (\Delta\omega_1 + \Delta\omega_2) (F_\psi + i \tau_\psi)],
    \label{eqCpsi}
\end{equation}
\begin{equation}
    C_\chi(\omega_1, \omega_2) \approx 2 A B [1 + (\Delta\omega_1 + \Delta\omega_2) (F_\chi + i \tau_\chi)],
    \label{eqCchi}
\end{equation}
with
\begin{align}
    \nonumber
    F_\psi = \frac{1}{|A|^4 + |B|^4 + 2 |A|^2 |B|^2 \cos(2 \delta)} \left\{ \left( \frac{|A|'}{|A|}|A|^4 + \frac{|B|'}{|B|}|B|^4 \right) + \right.\\
    \left. |A|^2|B|^2 \left[\cos(2\delta) \left(\frac{|A|'}{A} + \frac{|B|'}{B} \right) + \sin(2\delta) (\arg\{A\}' - \arg\{B\}') \right] \right\},
    \label{eqFpsi}
\end{align}
\begin{align}
    \nonumber
    \tau_\psi = \frac{1}{|A|^4 + |B|^4 + 2 |A|^2 |B|^2 \cos(2 \delta)} \left\{ \arg\{A\}' |A|^4 + \arg\{B\}' |B|^4 + \right.\\
    \left. |A|^2|B|^2 \left[\cos(2\delta) \left(\arg\{A\}' + \arg\{B\}' \right) + \sin(2\delta) \left( \frac{|B|'}{B} - \frac{|A|'}{|A|} \right) \right] \right\},
    \label{eqTpsi}
\end{align}
\begin{equation}
    F_\chi = \frac{1}{2} \left( \frac{|A|'}{|A|} + \frac{|B|'}{|B|} \right),
    \label{eqFchi}
\end{equation}
\begin{equation}
    \tau_\chi = \frac{1}{2} \left( \arg\{A\}' + \arg\{B\}'\right).
    \label{eqTchi}
\end{equation}
In these expressions $\arg\{A\}' = d\arg\{{\alpha}(\omega)\}/d\omega|_{\omega_{in}}$, $\arg\{B\}' = d\arg\{{\beta}(\omega)\}/d\omega|_{\omega_{in}}$, $\delta = \arg\{B\} - \arg\{A\}$, $|A|' = d|{\alpha}(\omega)|/d\omega|_{\omega_{in}}$, and $|B|' = d|{\beta}(\omega)|/d\omega|_{\omega_{in}}$. 

We further make the approximation $1 + x\Delta\omega \approx e^{x\Delta\omega}$ in Eqs. \eqref{eqCpsi} and \eqref{eqCchi}, which gives
\begin{equation}
    C_\psi(\omega_1, \omega_2) \approx (A^2 + B^2) \exp\left[(\Delta\omega_1 + \Delta\omega_2) (F_\psi + i \tau_\psi)\right],
    \label{eqCpsi_exp}
\end{equation}
\begin{equation}
    C_\chi(\omega_1, \omega_2) \approx 2 A B \exp\left[(\Delta\omega_1 + \Delta\omega_2) (F_\chi + i \tau_\chi)\right],
    \label{eqCchi_exp}
\end{equation}
and concludes the derivation of Eqs. (8) and (9) of the main text.

Further, Eqs. \eqref{eqCpsi_exp} and \eqref{eqCchi_exp} allow us to obtain the physical interpretation of the origin of the purity loss in the output state that we introduced in the main text. First, we inspect the two terms in the integral that gives $\ket{\Psi^o_+}$ (Eq. \eqref{eqPSI_OUT}),

\begin{align}
    \nonumber
    \phi(\omega_1, \omega_2) C_\psi(\omega_1, \omega_2) \approx\\
    A_\psi \exp\left[ -\frac{1}{2\sigma^2} (\omega_1 - (\omega_{in} - \Omega_\psi))^2  + i \omega_1 \tau_\psi\right] \exp\left[ -\frac{1}{2\sigma^2} (\omega_2 - (\omega_{in} - \Omega_\psi))^2  + i \omega_2 \tau_\psi\right],
    \label{eqPhiCpsi}
\end{align}
\begin{align}
    \nonumber
    \phi(\omega_1, \omega_2) C_\chi(\omega_1, \omega_2) \approx \\
    A_\chi \exp\left[ -\frac{1}{2\sigma^2} (\omega_1 - (\omega_{in} - \Omega_\chi))^2  + i \omega_1 \tau_\chi\right] \exp\left[ -\frac{1}{2\sigma^2} (\omega_2 - (\omega_{in} - \Omega_\chi))^2  + i \omega_2 \tau_\chi\right].
    \label{eqPhiCchi}
\end{align}
where $\Omega_\psi = \sigma^2 F_\psi$ and $\Omega_\chi = \sigma^2 F_\chi$. $A_\psi$ and $A_\chi$ are the same amplitudes that appear in Eqs. (8) and (9) of the main text and are given by 
\begin{equation}
    A_\psi = \frac{1}{\sigma \sqrt{\pi}} e^{-2\omega_{in}(i \tau_\psi + F_\psi + 2 \Omega_\psi) + 2\Omega_\psi^2} (A^2 + B^2)
\end{equation}
\begin{equation}
    A_\chi = \frac{2}{\sigma \sqrt{\pi}} e^{-2\omega_{in}(i \tau_\chi + F_\chi + 2 \Omega_\chi) + 2\Omega_\chi^2} A B
\end{equation}

Equations \eqref{eqPhiCpsi} and \eqref{eqPhiCchi} indicate that $\phi(\omega_1, \omega_2) C_\psi(\omega_1, \omega_2)$ and $\phi(\omega_1, \omega_2) C_\chi(\omega_1, \omega_2)$ describe two different Gaussian pulses. $\phi(\omega_1, \omega_2) C_\psi(\omega_1, \omega_2)$ has a central frequency $\Omega_\psi$ and central time $\tau_\psi$, while $\phi(\omega_1, \omega_2) C_\chi(\omega_1, \omega_2)$ has a central frequency $\Omega_\chi$ and central time $\tau_\chi$. Thus, the output state scattered by the nanoparticle is given, in this approximation, by a superposition of two contributions that are time delayed by $\Delta \tau = \tau_\Psi - \tau_\chi$ and frequency shifted by $\Delta \Omega = \Omega_\Psi - \Omega_\chi$. Note that this frequency shift is mediated by a reshaping effect due to the losses present in this system.

As indicated in Eq. (10) of the main text, either $\Delta \tau \neq 0$ or $\Delta \omega \neq 0$ (or both $\Delta \tau \neq 0$ and $\Delta \omega \neq 0$) introduce a loss of the purity of the output detected state. The differences between the output pulses are due to the frequency dependence of $\alpha(\omega)$ and $\beta(\omega)$. Moreover, if $\alpha(\omega)$ and $\beta(\omega)$ were constants in eqs. \eqref{eqAexp} and \eqref{eqBexp} then  $\tau_\psi$, $\tau_\chi$, $F_\psi$, and $F_\chi$ in eqs. \eqref{eqFpsi}-\eqref{eqTchi} would become zero. Thus, the helicity-splitting coefficients that show a fast spectral variation (fast compared with the spectral width of the incident pulse) lead to a significant loss of purity in the output post-selected state.

\subsection{Quantum response of a spherical nanoparticle that only supports electric or magnetic modes}
We show next that the loss of purity only occurs when the nanoparticle support both electric and magnetic resonances. A nanoparticle that only supports magnetic resonances satisfies $a_n = 0$ for all $n$ in eqs. \eqref{eqVn} and \eqref{eqWn}. In this case, following the derivation of section \ref{secS1} we obtain $\alpha(\omega) = -\beta(\omega)$ (see also chapter 4 of Ref. \cite{BohrenHuffman}, and Ref. \cite{Zambrana12}). On the other hand, if a nanoparticle only supports electric resonances, then $b_n = 0$ for all $n$ in eqs. \eqref{eqVn} and \eqref{eqWn}, and in this case we obtain $\alpha(\omega) = \beta(\omega)$. Both situations constitute a special case with $C_\chi(\omega_1, \omega_2) = \pm C_\psi(\omega_1, \omega_2) =\pm 2 \alpha(\omega_1)\alpha(\omega_2)$ in eqs. \eqref{eqCpsi} and \eqref{eqCchi} (the $+$ and $-$ sign corresponds to $\alpha(\omega) = + \beta(\omega)$ and $\alpha(\omega) = - \beta(\omega)$, respectively). Using eqs. \eqref{eqPSI_OUT}-\eqref{eqCChi0}, the output state in this case $\ket{\Psi_+^o}$ is
\begin{equation}
    \ket{\Psi_+^o} = \iint d\omega_1 d\omega_2 \phi(\omega_1, \omega_2) \left[ 2 \alpha(\omega_1)\alpha(\omega_2) \right]\left[ \ket{\psi_+^o(\omega_1, \omega_2)} \pm \ket{\chi_+^o(\omega_1, \omega_2)} \right]
    \label{eqPsiOEorM}
\end{equation}
In this expression, $\ket{\psi_+^o(\omega_1, \omega_2)}$ and $\ket{\chi_+^o(\omega_1, \omega_2)}$ are multiplied by the same spectral function $\phi(\omega_1, \omega_2) [2 \alpha(\omega_1)\alpha(\omega_2)]$. This shared weighting function causes the detected output state to be pure, as directly proved by inserting Eq. \eqref{eqPsiOEorM} in Eq. (5) of the main text. We obtain an output detected density matrix with only four non-zero terms, $\hat{\varrho}^o_{\psi_+, \psi_+} = \hat{\varrho}^o_{\chi_+, \chi_+} = 1/2$, $\hat{\varrho}^o_{\psi_+, \chi_+} = \hat{\varrho}^o_{\chi_+, \psi_+} = \pm 1/2$, which is an idempotent matrix, (i.e., $(\hat{\varrho}^o)^n = \hat{\varrho}^o$), and, thus, the output detected state is pure, $\text{Tr}\{(\hat{\varrho}^o)^2\} = 1$. 


\section{Loss of purity of all the states in the basis}

In the main text we considered the loss of purity of the incident state $\ket{\Psi^i_+} = \iint d\omega_1 d\omega_2 \phi(\omega_1, \omega_2) \ket{\psi^i_+(\omega_1, \omega_2)}$ (Eq. (1) of the main text). Next we consider the other input states that are frequency suppositions of the rest of the elements of the basis (Eqs. (1) and (2) of the main text),

\begin{subequations}
  \begin{tabularx}{\textwidth}{Xp{0cm}X}
  \begin{equation}
    \label{eqpsi-}
    \ket{\psi^{i}_{-}(\omega_1, \omega_2)} = \frac{1}{2} [\hat{a}^\dagger_i(\omega_1)\hat{a}^\dagger_i(\omega_2) - \hat{b}^\dagger_i(\omega_1)\hat{b}^\dagger_i(\omega_2)] \ket{0},
  \end{equation}
  \begin{equation}
    \label{eqchi-}
    \ket{\chi^{i}_{-}(\omega_1, \omega_2)} = \frac{1}{2} [\hat{a}^\dagger_i(\omega_1)\hat{b}^\dagger_i(\omega_2) - \hat{b}^\dagger_i(\omega_1)\hat{a}^\dagger_i(\omega_2)] \ket{0},
  \end{equation}
  \begin{equation}
    \label{eqchi+}
    \ket{\chi^{i}_{+}(\omega_1, \omega_2)} = \frac{1}{2} [\hat{a}^\dagger_i(\omega_1)\hat{b}^\dagger_i(\omega_2) + \hat{b}^\dagger_i(\omega_1)\hat{a}^\dagger_i(\omega_2)] \ket{0}.
  \end{equation}
  \end{tabularx}
\end{subequations}

First we consider the state
\begin{equation}
    \ket{X^{i}_{-}} = \iint d\omega_1 d\omega_2 \phi(\omega_1, \omega_2) \ket{\chi^{i}_{-}(\omega_1, \omega_2)}.
\label{eqCHI_M_IN}
\end{equation}
We consider in all of this work that the two-photon spectral amplitude is symmetric, \textit{i.e.} it does not change under the permutation of the arguments $\phi(\omega_1, \omega_2) = \phi(\omega_2, \omega_1)$. On the other hand, the state $\ket{\chi^{i}_{-}(\omega_1, \omega_2)}$ is antisymmetric under permutation of the frequencies. Thus, the incident state, $\ket{X^{i}_{-}}$, cannot be generated:
\begin{align}
    \ket{X^{i}_{-}} = \iint d\omega_1 d\omega_2 \phi(\omega_1, \omega_2) \frac{1}{2} [\hat{a}^\dagger_i(\omega_1)\hat{b}^\dagger_i(\omega_2) - \hat{b}^\dagger_i(\omega_1)\hat{a}^\dagger_i(\omega_2)] \ket{0} = 0.
\end{align}

Next, we consider the state
\begin{equation}
    \ket{\Psi^{i}_{-}} = \iint d\omega_1 d\omega_2 \phi(\omega_1, \omega_2) \ket{\psi^{i}_{-}(\omega_1, \omega_2)}
\label{eqPSI_M_IN}
\end{equation}
Using the transformation given in Eq. \eqref{eqTRANS} we obtain the output state scattered by the nanoparticle:
\begin{align}
    \nonumber
    \ket{\Psi^{o}_{-}} = \iint d\omega_1 d\omega_2 \phi(\omega_1, \omega_2) \left\{[{\alpha}(\omega_1){\alpha}(\omega_2)-{\beta}(\omega_1){\beta}(\omega_2)] \ket{\psi^{o}_{-}(\omega_1, \omega_2)} \right. +\\
    \left. [{\alpha}(\omega_1){\beta}(\omega_2)+{\beta}(\omega_1){\alpha}(\omega_2)] \ket{\chi^{o}_{-}(\omega_1, \omega_2)} \right\}.
\end{align}
Due to the use of indistinguishable photons, $\iint d\omega_1 d\omega_2 \phi(\omega_1, \omega_2) [{\alpha}(\omega_1){\beta}(\omega_2)+{\beta}(\omega_1){\alpha}(\omega_2)] \ket{\chi^{o}_{-}(\omega_1, \omega_2)} = 0$, and the only contribution to $\ket{\Psi^{o}_{-}}$ are the $\ket{\psi^{o}_{-}(\omega_1, \omega_2)}$ states,
\begin{align}
    \nonumber
    \ket{\Psi^{o}_{-}} = \iint d\omega_1 d\omega_2 \phi(\omega_1, \omega_2) [{\alpha}(\omega_1){\alpha}(\omega_2)-{\beta}(\omega_1){\beta}(\omega_2)] \ket{\psi^{o}_{-}(\omega_1, \omega_2)}.
\label{eqPSI_M_OUT}
\end{align}
This output state consists on only one state of the output two-photon basis, $\ket{\psi^{o}_{-}(\omega_1, \omega_2)}$. Thus, the output detected density matrix of this state (Eq. (5) of the main text) has only one element and it is pure. This could have been expected from the mirror and cylindrical symmetries of this particular state \cite{Buse18, Lasa20}.

Last, we consider the incident state

\begin{equation}
    \ket{X^{i}_{+}} = \iint d\omega_1 d\omega_2 \phi(\omega_1, \omega_2) \ket{\chi^{i}_{+}(\omega_1, \omega_2)},
\label{eqCHI_P_IN}
\end{equation}
and apply Eq. \eqref{eqTRANS}:
\begin{align}
    \nonumber
    \ket{X^{o}_{+}} = \iint d\omega_1 d\omega_2 \phi(\omega_1, \omega_2) \Big\{[{\alpha}(\omega_1){\alpha}(\omega_2)+{\beta}(\omega_1){\beta}(\omega_2)] \ket{\chi^{o}_{+}(\omega_1, \omega_2)} +\\
    \nonumber
    [{\alpha}(\omega_1){\beta}(\omega_2)+{\beta}(\omega_1){\alpha}(\omega_2)] \ket{\psi^{o}_{+}(\omega_1, \omega_2)} \Big\}=\\
    \left[C_\psi(\omega_1, \omega_2) \ket{\chi_{+}^{o}(\omega_1, \omega_2)} + C_\chi(\omega_1, \omega_2) \ket{\psi_{+}^{o}(\omega_1, \omega_2)} \right],
\end{align}
which is the same result obtained for the output state $\ket{\Psi^o_+}$ (Eq. \eqref{eqPSI_OUT}) except that the coefficients $C_\psi(\omega_1, \omega_2)$ and $C_\chi(\omega_1, \omega_2)$ of the states $\ket{\chi_{+}^{o}(\omega_1, \omega_2)}$ and $\ket{\psi_{+}^{o}(\omega_1, \omega_2)}$ are interchanged. By applying Eq. (5) of the main text we arrive to a similar output post-selected density matrix of that of the $\ket{\Psi^o_+}$ state. For the $\ket{X^{o}_{+}}$ state, the off-diagonal and the diagonal elements are interchanged, \textit{i.e.}, the $\hat{\varrho}^o_{\psi_+,\psi_+}$, $\hat{\varrho}^o_{\chi_+,\chi_+}$, $\hat{\varrho}^o_{\psi_+,\chi_+}$, and $\hat{\varrho}^o_{\chi_+,\psi_+}$ density matrix elements of the output state $\ket{\Psi_+^o}$ are equal to the $\hat{\varrho}^o_{\chi_+,\chi_+}$, $\hat{\varrho}^o_{\psi_+,\psi_+}$, $\hat{\varrho}^o_{\chi_+,\psi_+}$, and $\hat{\varrho}^o_{\psi_+,\chi_+}$ elements of the output state $\ket{X_+^o}$, respectively. However these changes do not affect the loss of purity, and thus, the output post-selected density matrix obtained for $\ket{X_+^o}$ has the same purity as the one studied in the main text for the incident state $\ket{\Psi_+^i}$.

\newpage
\bibliography{bibliographySI}
\bibliographystyle{ieeetr}